\documentstyle[prl,aps,epsfig]{revtex}
\begin{document}
\twocolumn[\hsize\textwidth\columnwidth\hsize\csname@twocolumnfalse%
\endcsname
\title{Response to Greiter's Comment} 
\author{Eugene Demler, Shou-Cheng Zhang}
\address{ Department of Physics, Stanford University, Stanford, CA
  94305} 
\author{Stefan Meixner and Werner Hanke}
\address{Institut f\"{u}r Theoretische Physik, Am Hubland,
D-97074 W\"{u}rzburg, Federal Republic of Germany}
\date{May 15, 1997}

\maketitle 

\begin{abstract}
  Greiter\cite{Greiter} claimed erroneously 
  that the $\pi$ excitation\cite{pi} of the Hubbard 
  model has an energy of the order of $U$.  This 
  mistake originates
  from a inconsistent treatment of the two particle Hartree energy 
  and the Hartee correction to the chemical potential.
  In any self-consistent calculations, these two contributions cancel
  exactly, as shown by Kohn, Luttinger and Ward\cite{KLW}.
  We also show that his interpretation of the finite-size studies
  is inconsistent.
\end{abstract}


]

In ref \cite{pi}, we studied the $\pi$ resonance within the 
Hubbard and the $t-J$ model. We calculated the spectrum of the 
$\pi$ operator 
\begin{eqnarray}
\pi_{d}^\dagger= \sum_{\bf{k}} (\cos k_x - \cos k_y) \;
      c_{{\bf k} + {\bf Q},\uparrow}^\dagger 
      c_{-{\bf k},\uparrow}^\dagger
      \  \ {\bf Q}=(\pi,\pi)
\label{pi}
\end{eqnarray}
within the $T$ matrix approximation (which is equivalent to the 
Hartree-Fock approximation in the two particle channel), and showed that
{\it within this approximation}, the spectrum consists of a sharp 
pole at \mbox{$\omega_0=-2\mu +J/2~ (1-n)$}, where $\mu$ is the
chemical potential approperiate for the 
t-J model. Meixner {\it et al.}\cite{Stefan1}
 have checked this result within 
exact numerical diagonalization, and found excellent agreement.

Greiter\cite{Greiter} claimed to have found an error in our Hartree-Fock 
calculation, and claimed to have shown that the energy spectrum of the $\pi$
operator is of the order of $U$, and it is incoherent. Both of these claims
are wrong, and we shall trace his mistake in this note.

Before proceeding to the calculations, let us first show the physical origin of
his error. The $\pi$ operator inserts two extra particles in triplet channel
into a fermi system. In the grand canonical ensemble, the energy of the 
$\pi$ operator is defined with respect to twice the chemical potential.
Since the $\pi$ operator is a spin triplet, the on-site Hubbard $U$ interaction can not
enter the multiple scattering process between these two extra
particles directly but only through a Heisenberg exchange term $J$.
Within the Cooper approximation, the only role of the fermi sea is to block 
the filled states in the multiple scattering process, and the chemical potential
is simply given by the fermi energy of the free fermi sea. At this
zeroth order, 
one can clearly see that the energy of the $\pi$ operator is
independent of $U$. 
To the first order in $U$, one can consider the interaction
between the inserted electron with the rest of the electrons inside the fermi
sea, and there is a non-zero Hartree contribution $2Un_\downarrow$. However,
to this order, there is also a Hartree contribution to the chemical potential.
Since the energy of inserting a single electron into the fermi sea is also 
raised by $Un_\downarrow$, these two contributions cancel each other exactly, 
and therefore the energy of the $\pi$ operator is unchanged with the inclusion
of the Hartree process. There is no Fock process for the Hubbard $U$.
Greiter reached the erroneous conclusion because he
only took the Hartree contribution to the two particle energy, while
overlooking the correction to the chemical potential which is of
the same order.

In a series of classic papers, Kohn and Luttinger, Luttinger and Ward\cite{KLW}
have shown that this type of cancelation occurs order by order in perturbation
theory. Therefore, one can make a stronger argument, namely the energy
of the $\pi$ resonance will remain finite in the large $U$ limit,
if the perturbation series converges in this limit. However, this assumption
may break down, and the Hartree treatment of the
interaction would certainly be inadequate. Unfortunately, no systematic 
calculations of the $\pi$ spectrum have been carried out in the limit
where double occupancy is strictly forbidden. Greiter's comment is
purely based 
on the Hartree-Fock approximation, and did not add anything to the analytical 
calculations in this limit. Meixner {\it et al}\cite{Stefan1} 
presented strong numerical evidence that the $\pi$ spectrum remains sharp 
in this limit, and have shown furthermore the low energy of the peak scales 
{\it inversely} with $U$. Greiter tried to argue that the $\pi$ peak
could be due to mixing with the spin correlation function. Close to
half-filling, there is considerable pairing fluctuation, and we indeed
first suggested that these two channels are mixed. If one only looks at
the correlation functions close to half filling, there is an obvious
``chicken and egg" problem in interpretation. However, Meixner {\it et al.}
found that the sharpness of the $\pi$ peak persists over a wide doping range,
up to $n=0.4$\cite{Stefan2}. In this doping regime, there is no sign of any pairing
fluctuations, and the sharpness of the $\pi$ peak must be intrinsic, 
completely decoupled from the spin correlation function. 

For most readers familiar with the basics of the Hartree-Fock or $T$ matrix
approximation, the above discussion should have adequately addressed the
elementary nature of the debate raised by Greiter and the source of his error.
However, in order to make this note self-contained, we shall show the details
of the calculations in the remainder of this note. 

The Hubbard model is given by
\begin{eqnarray}
{\cal H} = \sum_{k\sigma} \epsilon_{k\sigma} c_{k\sigma}^{\dagger}
c_{k\sigma} + U \sum_i n_{i\downarrow} n_{i\uparrow} - \mu \sum_i
n_{i\sigma} 
\end{eqnarray}
For a given density of electrons $n$, the chemical potential $\mu$ is 
implicitly determined from $n = -\partial E_0 / \partial \mu$, where 
$E_0$ is the ground state energy. Order by order in perturbation theory,
the dependence of $E_0$ on $\mu$ changes. In order to keep the density
of the system fixed at any order, one has to renormalize $\mu$ consistently.
In general, $\mu$ determined from the above relation changes from order to
order in perturbation theory. This point is discussed in detail by Kohn,
Luttinger and Ward\cite{KLW}.

Let us see how the Hartree-Fock approximation renormalizes the chemical
potential. For non-interaction electrons~ 
$n =  \frac{2}{N} \sum_k \theta(\mu -\epsilon_k )$.
This determines the chemical potential in zeroth order to be the
non-interacting fermi energy $\mu_0=\epsilon_F$. Next we can consider the
commutator between the Hamiltonian and the one electron creation operator,
evaluated within the Hartree-Fock approximation,
\begin{eqnarray}
[ H, c_{k\sigma}^{\dagger} ] =
( \epsilon_k - \mu  
 + U n_{-\sigma} ) c_{k\sigma}^{\dagger} 
\end{eqnarray}
To this order, the ground state wave function is modified to be
\begin{eqnarray} 
c_{k\sigma}^{\dagger} | \Omega \rangle = 0  
\ \ if\ \  
\epsilon_k - \mu + U n_{-\sigma} < 0 \nonumber\\
 c_{k\sigma} | \Omega \rangle = 0  
\ \ if\ \ 
\epsilon_k - \mu + U n_{-\sigma} > 0
\end{eqnarray}
In this new ground state, the density of electrons is given by 
\begin{eqnarray}
n=\frac{1}{N} \sum_{k\sigma} \langle \Omega | c_{k \sigma}^{\dagger}
c_{k \sigma} | \Omega \rangle = \frac{2}{N} \sum_k \theta( 
 \mu - U n_{\downarrow} -\epsilon_k)
\end{eqnarray}
Comparing this expression with the density of the non-interacting electron
gas, we see that in order to fix the density, the chemical potential at
this order is given by $\mu= U n_{\downarrow} +\epsilon_F$.

At the same order in $U$, the energy of the $\pi$ operator is determined 
by evaluating the commutator in the Hartree-Fock approximation,
\begin{eqnarray}
[ H , \pi_d^{\dagger} 
] = 2 ( U n_{\downarrow} - \mu ) \pi_d^{\dagger} +
(terms\ of\ the\ order\ of\ J)
\end{eqnarray}  
Here the terms of the order of $J$ arises from multiple scattering process
of the added particles with each other. In lowest order, the $J$ terms can only
enter if we add a Heisenberg exchange term to the Hubbard model. This term is
introduced in order to be able to show a simple explicit result within
the lowest order Hartree-Fock theory. Its introduction is not necessary if
one performs a full diagrammatic calculation of the Hubbard model, as we will 
show later in this note. Anyway, the effect of $J$ is not challenged by 
Greiter's comment and we will not discuss it explicitly.
Since the added electrons form a spin triplet, the
on-site $U$ does not contribute to their mutual interaction. 
The first term arises from the Hartree
interaction of the added electrons with the electrons inside the fermi
sea and represents a self-energy correction to a one electron's Green's
function.
At this point, Greiter argued incorrectly that the energy of the $\pi$
operator is of the order of $U$. Because of the renormalization of the
chemical potential discussed in the previous paragraph, we see that the
$U$ term cancel and $2 ( U n_{\downarrow} - \mu )$ simply reduces to
$-2\epsilon_F$, which is manifestly independent of $U$.

The same effect can of course also be addressed directly in the canonical
formalism, without introducing the chemical potential. In this approach,
the energy of the $\pi$ operator is given by
\begin{eqnarray}
\omega_0 = [E_\pi(N+2)-E_0(N)] - 2 [E_0(N+1)-E_0(N)]
\label{canonical}
\end{eqnarray}
where $E_\pi(N+2)$ is the energy of a lowest excited state of the
$N+2$ particle  
system with total spin one and momentum {\bf Q}. The numerical calculations
of Meixner {\it et al.}\cite{Stefan1} uses this definition for 
the energy of the $\pi$ operator. If we neglect the
interaction between the added electrons with the background, the first
term is of the order of $J$\footnote{ For the tight binding
  Hamiltonians the dispersion is given 
  by $\epsilon_p = -2 t (cos p_x + cos p_y )$, and adding two particles
  with a center of mass momentum $Q=(\pi,\pi)$ results in a zero kinetic
  energy $\epsilon_p + \epsilon_{-p+Q} = 0$. The first term in
  (\ref{canonical}) is therefore determined by a repulsive interaction
  of two particles in a triplet state at nearest sites }
  and the second term is simply $-2\epsilon_F$.
If we include the Hartree interaction with the background, both terms
are raised by $2Un_{\downarrow}$, and cancel each other. We obtain the
same expression as before.

The reader may wonder if this type of cancelation happens only accidentally
in lowest order of perturbation theory. In fact, by using a powerful
theorem due to Luttinger and Ward\cite{KLW}, one can show that it occurs at every order
in perturbation theory. The energy of the $\pi$ resonance within the pure
Hubbard model can be obtained
by solving the following exact integral equation
\begin{eqnarray}
&&\Gamma_{pp'}(q,\omega) = \Gamma_{pp'}^{irr}(q,\omega)+  \\
& \int & \frac{d\omega'' d^2 p''}{(2 \pi)^3}
\Gamma_{pp''}^{irr}(q,\omega)
G(p'',\omega'') G(q-p'',\omega-\omega'') \Gamma_{p''p'}(q,\omega)
\nonumber
\end{eqnarray}
Here $\Gamma^{irr}$ is the sum of all two particle graphs which do not
fall into two separate pieces when one cuts two Greens function lines, and $G$
is the exact Greens function given by
\begin{eqnarray}
G(\omega,p) = \frac {1}{\omega-\epsilon_p-\mu-\Sigma(\omega,p)}
\end{eqnarray}
and $\Sigma(\omega,p)$ is the exact self-energy.  
Because of the spin triplet nature of the $\pi$ resonance, the symmetry of the
vertex function exactly projects out the on-site interaction part in
$\Gamma^{irr}$. The interaction on the nearest neighbor sites due to the remaining part 
in $\Gamma^{irr}$ can be {\it defined} to be the effective spin exchange
constant $J$. 
 $\Sigma(\omega,p)$ can involve many corrections, including the
on-site interaction, just as we have seen in the first order calculation. 
However, the chemical potential also receives a similar correction. 
Luttinger and Ward\cite{KLW} obtained an explicit formula relating these corrections,
\begin{eqnarray}
\epsilon_F = \mu - \Sigma(\mu,p_F)
\end{eqnarray}
which is valid to all orders of perturbation theory.
We can identify the constant part of $\Sigma(\omega,p)$ with the corrections due to an 
effective instantaneous on-site interaction. From the Luttinger Ward formula, we
see that any such correction is exactly canceled by a corresponding shift 
in the chemical potential. The remaining part of the self-energy renormalizes
the bare electron into a dressed electron with modified fermi parameters. It is 
plausible that in the low energy sector, the above integral equation has a
form similar to the lowest order Hartree-Fock equation and a low
energy collective pole exists as its solution. 

In conclusion we have identified the error in Greiter's comment: it
results from 
a inconsistent treatment of the Hartree correction to the two particle
energy and 
the corresponding correction to the chemical potential. We
demonstrated explicitly the  
cancelation of these two types of corrections within the lowest order
Hartree-Fock or the  
$T$ matrix calculation. Using a theorem due to Luttinger and Ward, we
argued that 
this type of cancellation occurs to all orders of perturbation theory.
Greiter's interpretation of the numerical results overlook a large data set 
on the $\pi$ resonance. 
In the parameter regime where pairing correlations are present, we
argued previously 
that the spin correlation and the $\pi$ correlation may mix. However,
$\pi$ resonance  
has been clearly identified numerically in the regime of low electron
density where 
the pairing correlation is absent, and is therefore intrinsic.

This work is supported by the NSF under grant numbers DMR-9400372 and
DMR-9522915.

\end{document}